# First-principles studies of oxygen interstitial dopants in RbPbI$_3$ halide for perovskite solar cells

Chongyao Yang[1], Wei Wu[1,2]†, Kwang-Leong Choy[1,3]*

[1]UCL Institute for Materials Discovery, University College London, Malet Place, London WC1E 7JE, United Kingdom
[2]UCL Department of Physics and Astronomy, University College London, Gower Street, London WC1E 6BT, United Kingdom
[3]Division of Natural and Applied Sciences, Duke Kunshan University, Kunshan, Suzhou, Jiangsu Province, China 215316

†Corresponding email: wei.wu@ucl.ac.uk
*Corresponding email: kwang.choy@duke.edu

## Abstract

Recent research on perovskite solar cells has caught much attention for the application in renewable energy materials. However, the effect of external dopants on the performance of perovskite solar cell is yet to be understood properly. Oxygen atom or molecule is important dopant to influence the stability of structural, electronic and optical properties as well as the performance of perovskite solar cells. RbPbX$_3$-type perovskites have fantastic chemical stability and good power conversion efficiency. Here for the first time, we have studied the effect of interstitial oxygen atom (O$_1$) and molecule (O$_2$) on the structural properties, and hence the electronic structure of RbPbI$_3$ from first principles. A significant reduction of the band gap from ~2.6 eV to ~ 1.0 eV, which is close to the optimal band gap according to the Shockley–Queisser limit, has been predicted when incorporating oxygen. This could in turn be applied to improve the optical properties for harvesting light if we can control the oxygen level appropriately. In addition, an exotic metallic state has been found in our calculations for interstitial oxygen molecule when there are strong O-O,

O-Pb, and O-I bonds, indicating the complex nature of oxygen-doped perovskite solar cells. The comparison between oxygen atom and molecules is consistent with the previous report about oxygen-molecule passivation of perovskite solar cells. This indicated oxygen incorporation can not only improve efficiency and stability but also facilitate the optimal band-gap engineering. Our work has therefore provided an important and timely theoretical insight to the effect of oxygen dopants in perovskite solar cells. Moreover, these results also provide theoretical foundation for further simulations such as molecular dynamics.

## I.  Introduction

Solar cells convert sunlight into electricity through the photovoltaic effect, without directly generating greenhouse gases such as carbon dioxides[1]. Solar energy has a huge potential to meet the future energy consumption[1,2]. Moreover, fossil fuel energy is non-renewable; economists predict that by 2040 the price of fossil fuel will triple that of 2010 due to shortage[3]. Therefore, solar energy resources, which can potentially preserve our natural environment, are extremely promising in the future energy market. There are so far three generations of solar cells. Silicon, which has a high power conversion efficiency (PCE) of 25% despite a high cost of ~0.3 $/W, was used for the first-generation photovoltaic technology (PV) [4,5]. Thin-film solar cells are the second generation, which can be prepared based on a thin film made of plastic, glass, or metal raw materials[6]. The main advantage of thin-film solar cell is its lower fabrication cost as compared with silicon counterpart. Flexibility is another advantage for thin-film solar cells, which implies the device can be fabricated using printing technologies and applied on surfaces. However, low PCE and instability (as compared with silicon) limits the application of thin films technologies to solar cells. The third-generation solar cell was born in the emerging solution-processed multi-layer cell structure. They represent the most advanced solar power generation technologies nowadays, with four branches: quantum-dot solar cells[1,7,8], organic solar cell[9], dye-sensitized solar cell (DSSC)[10], and perovskite solar cell (PSC)[1]. Among these, PSC is particularly promising because it could combine the strengths of the previous generations, including advanced thin film technology, mechanical flexibility, low cost, and efficient production[11]. PSC consists of a perovskite-structure

semiconductor as light-absorbing material, which is very suitable for thin film solar cells[1]. People can also integrate inorganic and organic materials in PSC to optimize solar cell performance. The crystal structure of perovskite can usually be expressed as $ABX_3$ with A and B sites for cations and X for anion. A can be organic cations like $CH_3NH_3^+$ or inorganic cations such as $Cs^+$. B usually represents divalent metal cations such as $Pb^{2+}$ or $Sn^{2+}$. X usually represents halide anions such as $I^-$, $Br^-$ and $Cl^-$. Due to the good solubility of perovskite[12], many low-cost and highly efficient processing methods can be used to fabricate PSC thin films, such as spin coating[13], screen printing[14], dip coating[14] and spray coating[15]. The performance of organic-inorganic mixed perovskite is remarkable because the combination of organic and inorganic materials results in a large binding energy between electron and hole, leading to strong photoluminescence and tunable conductivity[16][17][18]. Due to the low cost, excellent optical and electrical properties, PSC has become one of the research focuses for next generation solar cells. The PCE of PSC was only 3.8% in 2009, which has been improved to 19% in only three years[19]. The latest experimental results suggested the PCE of PSC has exceeded 25%[20][21].

Oxygen is an important species affecting the stability and performance of PSC[22][23]. Current research shows that oxygen has both negative and positive effects on hybrid organic-inorganic halide perovskite (HHPs). On the one hand, oxygen can form chemical bonds with organic cation, which makes PSC decompose faster. On the other hand, if oxygen can occupy halide vacancies on the surface of HHPs, passivation would occur, thus improving photoluminescence intensity[24]. To study the effect of oxygen dopants in PSC, inorganic halide perovskite (IHPs) could be a good choice because this can isolate the interaction between oxygen and inorganic part of PSC. Among the IHPs, the most promising candidates include $CsPbBr_3$, $CsPbI_3$, $CsPbIBr_2$, $CsPbI_2Br$, and $RbPbI_3$ due to their superb structural stability. Previously $CsPbBr_3$ has been studied experimentally to improve the material efficiency and stability. Screen printing technology has used to fabricate solar cells based on $CsPbBr_3$ to improve open circuit voltage[25]. A mechanochemical green-chemistry method without solvent has been implemented to grow $CsPbBr_3$ with a reduced band gap ~2.22 eV[26]. The B- and A- site cation doping in $CsPbBr_3$ has also been reviewed for the purposes of stability, luminescence performance, and spontaneous strains[27]. The effect of oxygen dopant on $CsPbI_2Br$ has been studied previously, which has shown that oxygen atoms can passivate CsPbI$_2$Br, leading to higher PCE and better stability[28]. Previously the size dependence of the lattice constant and band gaps for CsPbI$_3$ nanocrystals has been discussed[29]. It is worth noting that RbPbX$_3$ perovskite such RbPbI$_3$ is even more stable than CsPbI$_3$[30]. CsPbI$_3$ will undergo a reversible phase transition of octahedral structure to cubic structure at a temperature of ~600 K, while RbPbX$_3$ can maintain the phase stability at this temperature. However, the band gap for RbPbI$_3$ and CsPBI$_3$ remain rather large (~ 2.6 eV), which may hinder their applications to solar cells according to the Shockley-Queisser limit[31].

Previously density functional theory (DFT) was used to calculate the electronic structure and band gaps of CsPbI$_3$ and RbPbI$_3$[30]. Classical molecular dynamics simulations and first principles calculations have been performed to study the stability and phase transformation in CsPbI$_3$[32]. Recent crystal structure refinement of black-phase CsPbI$_3$ and corresponding band structure calculations suggest the material could possess an orthorhombic structure rather than cubic as conventionally adopted[33]. The phase and structural stabilities and the improving strategy for CsPbI$_3$ have also been reviewed both from the perspectives of experiment and theory[34]. A variety of density functionals, including Perdew-Bourke-Ernzerhof (PBE)[35] and screened hybrid-exchange function (HSE06[36]), have been chosen, in combination with spin-orbit interaction, to compute the electronic structure of CsPbI$_3$ and RbPbX$_3$. This can provide information about the band structure and electric (optoelectronic) properties of CsPbI$_3$ and RbPbX$_3$, which can lay a theoretical foundation for the follow-up research. Previously the electronic structure of pristine RbPbI$_3$ has been computed from first principles[37]. To our best knowledge, the theoretical studies of dopants such as oxygen in RbPbI$_3$ is still absent. In this report, we have studied the effects of oxygen interstitial dopants (O atom and O$_2$) on the electronic structure of δ-RbPbI$_3$ (room temperature orthorhombic phase), which suggests oxygen bulk doping, especially oxygen single atom, could facilitate the improvement of solar cell performance due to reduced band gap. The remaining discussion falls into three sections. In the section II, we will discuss the computational details. In the section III, we will present our computational results. In the section IV, we will draw the general conclusions.

## II. Computational details

The electronic structure, including band structures, density of states and wave functions of pristine and doped $RbPbI_3$ (Figure 1) were computed by using self-consistent-field (SCF) DFT implemented in CRYSTAL17[38]. The band structure of $RbPbI_3$ has been plotted along the path $\Gamma$ (0,0,0) - $X$ (1/2,0,0) - $R$ (1/2,1/2,1/2) - $M$ (1/2,0,1/2) - $Y$ (0,1/2,0) - $\Gamma$ in the first Brilloin zone. GGA and hybrid-exchange density functionals for the exchange-correlation energy were used for comparison, including PBE, PBE0[39], and B3LYP[40]. Among these functionals, PBE takes into account the gradient of the charge densities as a variable to improve the functional, leading to so-called generalized gradient approximation. PBE0 and B3LYP combine GGA functional with Hartree-Fock exact exchange to balance the localization and delocalization of the electron wave functions. By comparing with the previous experimental work on the band gap (~ 2.64 eV)[41,42], the band gap of the pristine $RbPbI_3$ computed by PBE was more consistent with the experiments than those computed by PBE0 and B3LYP (not shown here), thus PBE functional has been chosen for all our calculations. The Van der Waals (vdW) forces have been taken into account by using the approximation provided by Ref.[43], and the parameters for the vdW forces for all the elements have been taken from Ref.[44]. The perovskite room-temperature orthorhombic crystal structure $\delta$-$RbPbI_3$ [45] belongs to the *Pnma* space group, with $a$ = 10.276 Å, $b$ = 4.779 Å and $c$ = 17.393 Å (the starting point of our calculations). The self-consistent convergence criteria for total energy was set to $10^{-6}$ Hartree, and an 8×8×8 Brillouin zone *k*-point grid was used. We have chosen the basis set and the effective core potential for Rb as suggested by the previous work on $RbNbO_3$[46]. Pb basis set and the corresponding effective core potential for perovskite oxides have been selected by similar methods[47]. For iodine, we have chosen the basis set and effective core potential for its anion in halides[48]. The 8-411d11G basis set for oxygen has been chosen[49]. The truncation of the Coulomb and exchange series in direct space is controlled by setting the Gaussian overlap tolerance criteria to $10^{-6}, 10^{-6}, 10^{-6}, 10^{-6}$, and $10^{-12}$. To accelerate SCF convergence, all the calculations have been performed adopting a linear mixing of Fock matrices by 30%. In our calculations, we have positioned $O_1$/$O_2$ near Rb, Pb, and I at the beginning of our calculations to see the effect of O on different element, and then optimized the crystal structure (both unit cell parameters and atomic positions) driven by the SCF process (Figure 1). Based on the optimized geometry, we have further performed single-point calculations and obtained the band structure and density of states (DOS), which can then be used to assess the effects of O dopants. The projected DOS (PDOS) is for all the atoms of the same species in the unit cell. The total DOS is in purple, the PDOS for Rb is in blue, Pb is in orange, I is in green, and O is in red, throughout the paper.

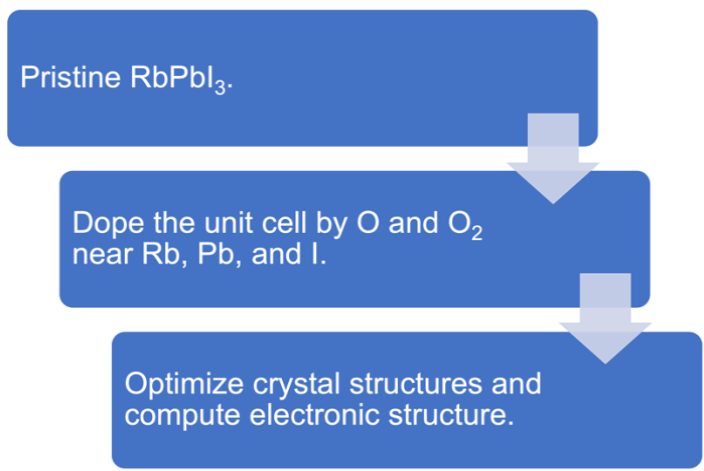

Figure 1: A flowchart for the modelling process. (a) We have first performed DFT calculations with different types of functionals to find the band structure of the pristine $RbPbI_3$ and compare them with experiment. According to the comparison with the experimental band gap, we have chosen the PBE functional. (b) Then we doped $O_1$/$O_2$ near Rb, Pb, and I to establish the understanding of their effects. (c) Based on the optimized crystal structures, we have obtained the band structure and DOS, which are useful for assessing the material properties.

## III. Results and Discussions

As shown in Figure 2, we have computed the band structure and DOS for pristine $RbPbI_3$. The unit cell parameter and atomic positions therein have been optimized with $a$ = 10.20 Å, $b$ = 4.33 Å, and $c$ = 17.46 Å. The relative difference between the optimized unit cell parameters and the experimental[45] are 0.8%, 10.2%, and 0.4% for $a$, $b$, and $c$, respectively. The previous experiments showed the band gap of $RbPbI_3$ was 2.64 eV [42]. The results from DFT calculations based on PBE were 2.84 eV without vdW forces (not shown) and 2.69 eV with vdW forces. Both calculation results are very close to the experimental data, which suggests the reliability of DFT method with PBE functional for δ-RbPbI$_3$. Moreover, the calculation with the vdW forces is closer to the experimental data, which implies that vdW forces have a significant impact on the band structure. Our calculation is also consistent with the previous first principles calculations using plane waves that show the band gap is ~ 2.5 eV[37]. We have also performed first principles calculations using PBE0 and B3LYP (not shown), both of which have provided far too large band gaps (> 4 eV). We have therefore chosen PBE functional + vdW forces to perform all the remaining calculations. In addition, from DOS and PDOS in Figure 2, we can see Pb is dominant on the conduction bands, I on the valence bands, while Rb has negligible contributions to the band structures near the Fermi energy, as expected. These results are also in agreement with the previous calculations reported in Ref.[42].

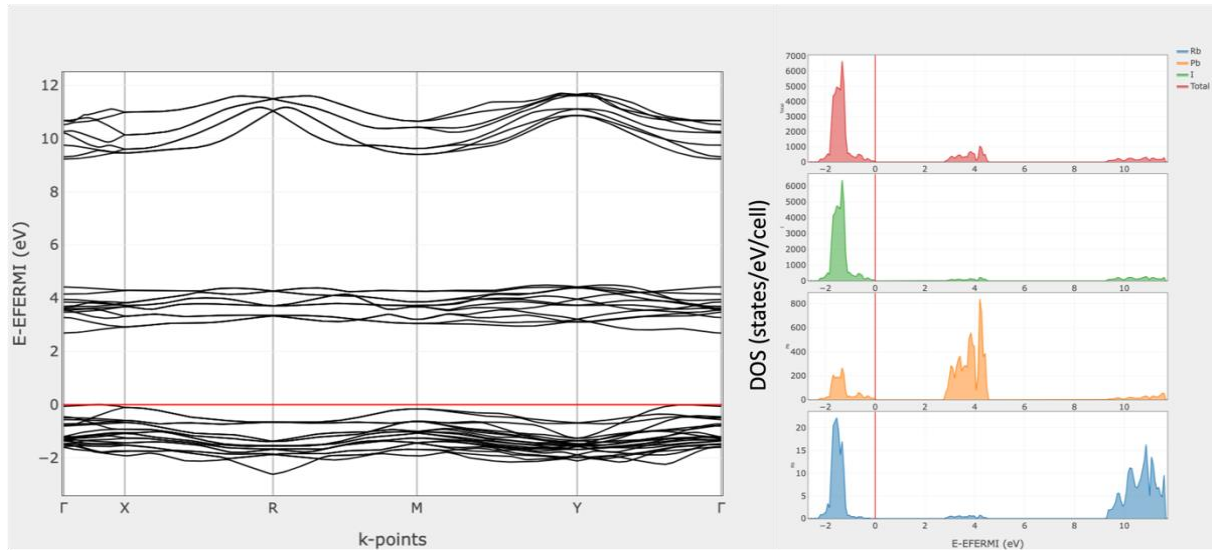

Figure 2: Band structure of pristine RbPbI$_3$ (left), and the corresponding DOS (right). The total DOS is in red. The PDOS for Rb is in blue, Pb in orange, and I in green. The red line indicates the Fermi level.

Subsequently, the effect of an oxygen interstitial dopant (O$_1$) added into the unit cell was investigated. An oxygen atom was inserted in the vicinity of Rb, Pb and I element, respectively. It is worth noting that the inserted atom was only near one atom and far away from others to ensure the consistency of the calculation results. However, the final optimized geometry would be determined by the self-consistent process. Without vdW forces, the computed band gap was 1.56 eV when adding an oxygen atom near Rb, whereas the bandgap was 1.22 eV when adding an oxygen atom near Pb, and 1.60 eV when adding an oxygen atom near iodine. All the results for the band gaps were less than 2 eV due to the formation of the dopant level within the band gap of pristine RbPbI$_3$ owing to oxygen and iodine atoms. Compared with the band gap of the pristine structure without vdW forces, after the oxygen atom was inserted, the band gap become smaller no matter which atom the oxygen was close to. In the calculation with vdW force, the band gap changes to 1.26 eV after adding an oxygen atom near Rb, 1.17 eV near Pb, and 0.91 eV near iodine. It is clear that the band gaps become even smaller than those calculations without the vdW forces, due to the dopant level. As shown in Figure 3, the total DOS (purple), O-PDOS (red), I-PDOS (green), Pb-PDOS (orange) and Rb-PDOS (blue) from top to bottom, have been computed for the situation where oxygen is close to Rb. Notice that all these band gaps with O$_1$ interstitial are near the optimal band gap for solar cell[31]. The other scenarios, i.e. the added oxygen was close to Pb and I (not shown here) share the similar qualitative feature, i.e., the formation of oxygen defect band within the pristine band gap. The oxygen and iodine atoms played a leading role in the defect band within the band gap. On the other hand, when the oxygen atom appears close to Rb and Pb (not shown), the peak of the oxygen atom projection would become higher and narrower, which suggested that the electron could be trapped, affecting the PCE due to the decreased electron or hole mobility. Therefore, it could be speculated that when the oxygen atom entering the bulk, it would form chemical bonds with iodine, thus influencing the band structure. Through a lateral comparison of the computed band gaps as shown in Table 1, it is found that the band gap decreases in both cases - with and without vdW force, after adding an oxygen atom. All the calculations suggest that in the optimized structure O atom will finally be rather close to I atom with a bond length ~ 2 Å, suggesting the importance of iodine for the geometry when incorporating O atom in the structure.

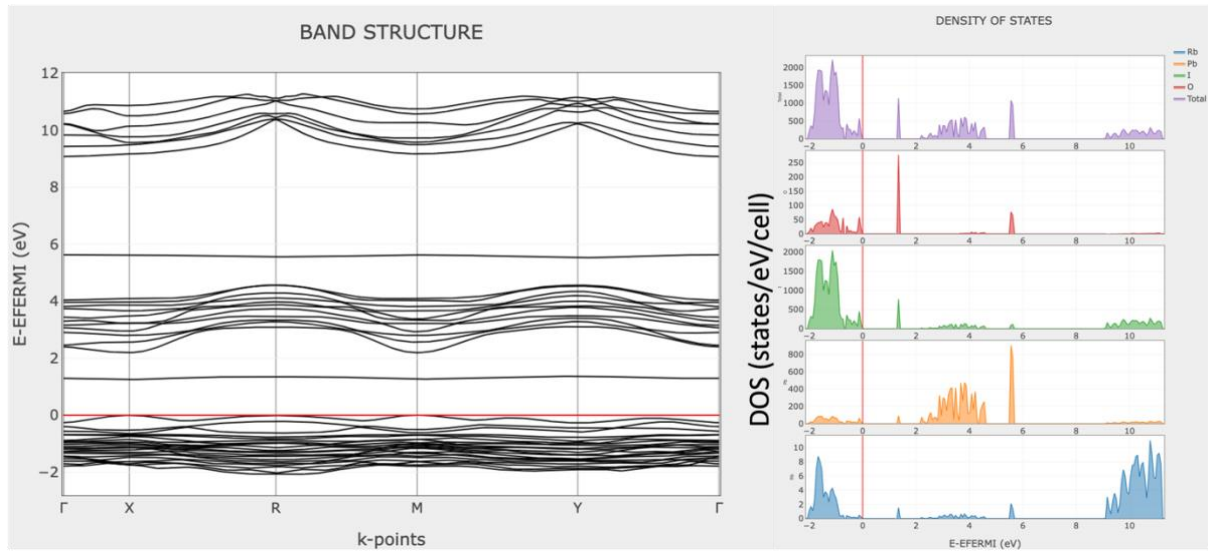

Figure 3: Band structure of $O_1$ interstitial near Rb (left), and the corresponding DOS (right) with vdW force. The total DOS is in purple. The PDOS for Rb is in blue, Pb in orange, I in green, and O in red. The red line indicates the Fermi level.

Furthermore, the influence of oxygen molecules ($O_2$) on the properties of $RbPbI_3$ was studied by inserting an oxygen molecule close to Rb, Pb (not shown here), and I, respectively. We have computed the electronic structures with and without vdW; in this report we only show the former scenario. Without vdW forces, when adding oxygen molecules near Pb and I, the band gaps are 1.03 eV and 0.72 eV, respectively, while the near-Rb calculation indicated a metallic state. It is obvious that the band gaps have been reduced to ~1 eV after adding oxygen molecules. Similar to $O_1$ interstitial, oxygen and iodine had evident dominance for the defect level. When oxygen molecules appeared around Pb (not shown here) and I, the peak of oxygen atom projection was high and narrow, implying strong electron trapping on O atoms.

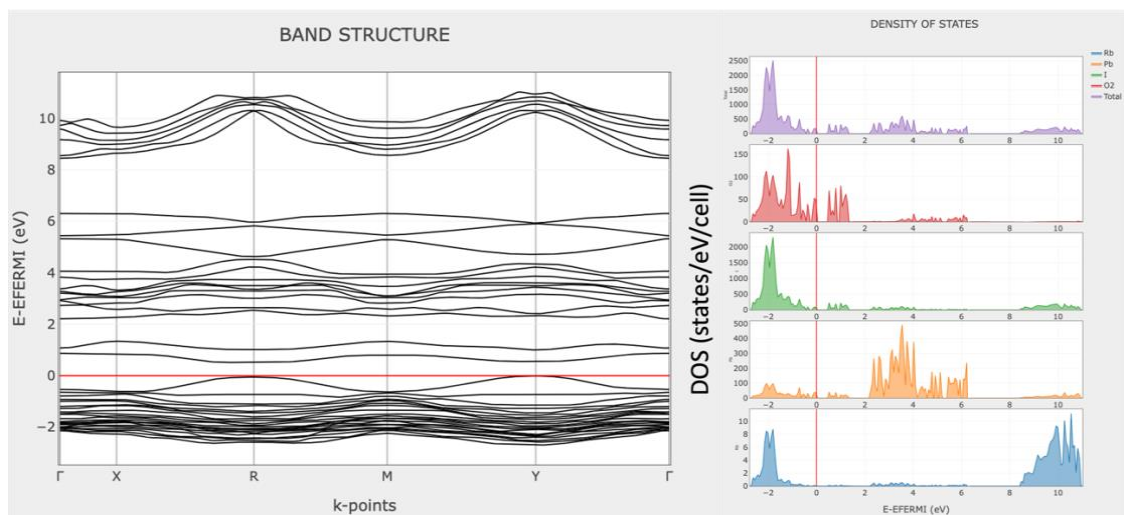

Figure 4: Band structure of $O_2$ interstitial near iodine, and the corresponding density of states with vdW force. The total DOS is in purple. The PDOS for Rb is in blue, Pb in orange, I in green, and O in red. The red line indicates the Fermi level.

From the band structure calculated with the vdW force, the band gap is 0.52 eV when adding an oxygen molecule near iodine, as shown in Figure 4, and 0.87 eV when adding an oxygen molecule near Pb (not shown), while adding an oxygen molecule near Rb will lead to a metallic state (Figure 5), consistent with the calculation without vdW forces. The band gap decreases even more than those in the single-oxygen calculations. It is worth mentioning that it becomes a metallic state after adding an oxygen molecule near Rb in the initial geometry, with a large band dispersion, in which iodine and oxygen atoms would make important contribution, as suggested in the PDOS. The order of the five PDOS is listed as the total (purple), O (red), I (green), Pb (yellow) and Rb (blue) from above to the bottom in Figure 4 and Figure 5. Moreover, oxygen and iodine dominated the DOS for the defect band within the band gap, which is distinguished from the pristine one.

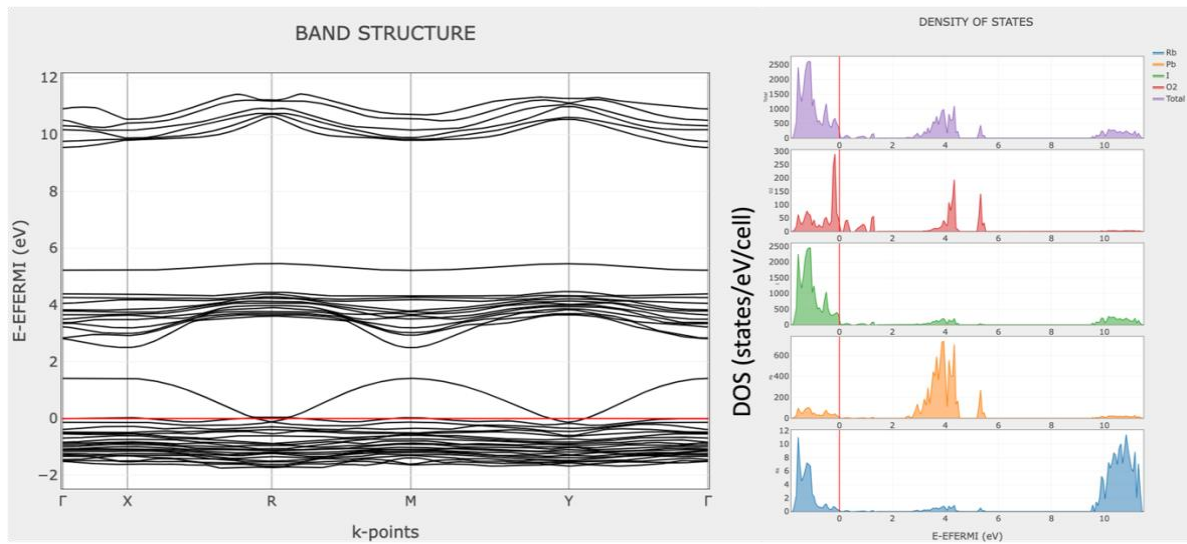

Figure 5: Band structure of $O_2$ interstitial near Rb, and the corresponding density of states with vdW force. The total DOS is in purple. The PDOS for Rb is in blue, Pb in orange, I in green, and O in red. The red line indicates the Fermi level.

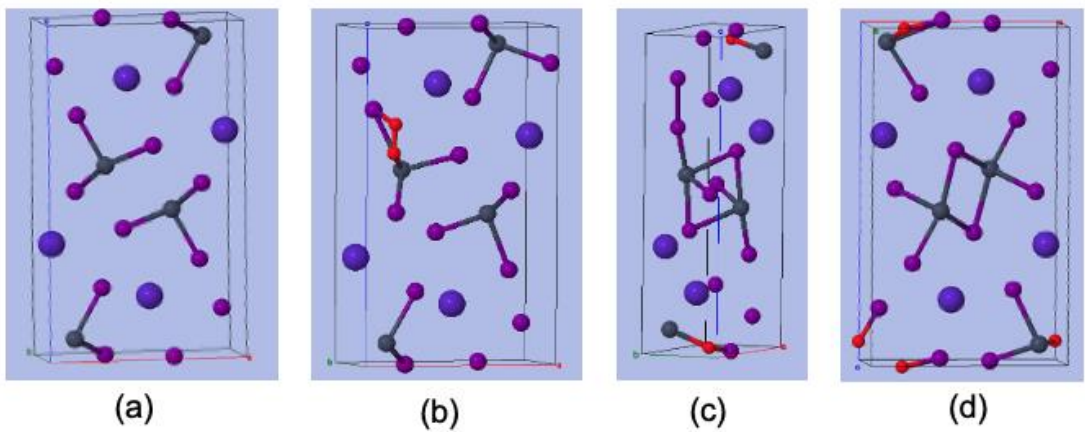

Figure 6: A comparison of the optimized geometries for the $O_2$ interstitial with vdW forces. Rb is represented by deep purple balls, Pb by dark grey, I by deep pink, and O in red. (a) the pristine structure, (b) the initial geometry where $O_2$ is near Rb, (c) near Pb, and (d) near I. The O-O bond length for (b) is ~1.5 angstroms, whereas it is ~ 3 angstroms for (c) and (d).

As shown in Figure 6, we can compare the optimized geometries in the unit cell for the four different situations: pristine, near-Rb, near-Pb, and near-I. The O-O bond length for near-Rb calculation is ~ 1.5 angstroms, which is slightly larger than that in $O_2$ molecule (1.2 Å). In addition, they are also bonded with the nearby Pb and I atoms (the shortest Pb-O bond length ~ 3.3 Å and I-O bond length ~2.5 Å). The O-O bond length in near-Rb situation is significantly smaller than that in the other cases, as shown in Figure 6. This indicates the physical origin of the metallic state for the near-Rb calculation might mainly be due to the strong O-O bond as it is well known that $O_2$ molecule is triplet diradical and could be highly reactive in a lattice environment despite the slightly longer bond length. Moreover, the high density of $O_2$ (in the order of $10^{21}$ per $cm^3$) might be another reason for the metallic state. By contrast, in the near-Pb and near-I calculations, the oxygen molecules are broken into two well separated oxygen atoms (O-O bond length ~ 3 angstroms), hence making smaller effect on the band gap, which is similar to the $O_1$ interstitial calculations except the weak interaction between oxygen atoms.

Table 1: A summary of band gap after adding oxygen atoms and molecules to $RbPbI_3$, with and without vdW force.

| Band gap (eV) | Pristine | $O_1$ near Rb | $O_1$ near Pb | $O_1$ near I |
|---|---|---|---|---|
| Without vdW force | 2.84 | 1.56 | 1.22 | 1.60 |
| With vdW force | 2.69 | 1.26 | 1.17 | 0.91 |
| Band gap (eV) | Pristine | $O_2$ near Rb | $O_2$ near Pb | $O_2$ near I |
| Without vdW force | 2.84 | Metallic | 1.03 | 0.72 |
| With vdW force | 2.69 | Metallic | 0.87 | 0.52 |

Table 1 shows the band gap values are reduced in both cases with and without vdW force after adding oxygen molecules. The calculated band gap with vdW force is smaller than that without vdW force. Oxygen atoms could provide empty band to accommodate excited electrons for inter-band transitions as shown in the band structure, which should have an important effect on PSC. It is evident that the band gaps after adding oxygen molecules become smaller than that with oxygen atoms in each condition. All the band gaps computed here are less than 2 eV after adding oxygen atom and oxygen molecules except the $O_2$ interstitial near-Rb scenario. Since the optimal band gap for solar cell material is approximately 1 eV according to the Shockley-Queisser limite[31], the performance of PSC could be improved provided we can control the oxygen level appropriately. As shown in Table 1, the calculations for all the $O_1$ interstitial scenarios and O2 interstitial near-Pb have an appropriate band gap (0.87 - 1.26 eV) to achieve a good PCE. The other calculations either indicate a small band gap (~0.5 eV) or metallic state, which is not desirable for solar cells. However, the peak of oxygen projection is very narrow, which might imply electrons can easily be trapped, thus decreasing the mobility of the electrons.

As shown in Table 2, we have analysed all the shortest Pb-O, I-O, Rb-O, and O-O bond lengths for the calculations with vdW forces. For $O_2$ interstitial, we have taken the average value of the two oxygen atoms. Comparing $O_1$ and $O_2$ calculations of the bond lengths, we can find that the electronic state will still be insulating if O is close to Pb and I (~ 2 Å). Otherwise, as in the case O2 interstitial near-Rb calculation, the two oxygen atoms can form strong bonds (~1.5 Å), which will lead to an exotic metallic state. In addition, the movement trend of oxygen atoms was found. When oxygen atoms or oxygen molecules were added to the bulk, they tend to move closer to Pb and I, thus forming Pb-O-I bonds, reducing the band gap or leading to a metallic state (if two oxygen atoms are close).

Table 2: Bond length analysis for the calculations with vdW forces. The shortest bonds have been taken into account. For the Pb-O, I-O, and Rb-O bonds for $O_2$ interstitial, we have taken the average of the two oxygen atoms.

| Calculations<br><br>Bond length (Å) | $O_1$ interstitial | | | $O_2$ interstitial | | |
|---|---|---|---|---|---|---|
| | Near Rb | Near Pb | Near I | Near Rb | Near Pb | Near I |
| Pb-O | 2.2 | 2.4 | 2.3 | 3.6 | 2.5 | 2.4 |
| I-O | 2.1 | 2.1 | 2.1 | 2.9 | 2.2 | 2.2 |
| Rb-O | 3.0 | 5.0 | 4.7 | 3.0 | 5.3 | 4.9 |
| O-O | / | / | / | 1.5 | 3.0 | 3.1 |

To understand the stability of PSC after oxygen interstitial, the defect formation energies of $O_1$ and $O_2$ interstitial dopants, with and without vdW force, are shown in Figure 7. The defect formation energies are computed as the energy difference between the optimized structure with dopants and the total energy of the optimized pristine structure and $O_1$/$O_2$ alone, which reads $E_D = E_{Doped} - E_P - E_{O_1/O_2}$, in which $E_{Doped}$ is the total energy of doped system, $E_P$ is the total energy of pristine system, $E_{O_1/O_2}$ is the total energy for $O_1$/$O_2$ alone. The blue and red curves are obtained by calculating without vdW force. The grey and yellow curves represent data with vdW force. It is clear that the absolute value of the defect formation energy of $O_2$ interstitial is smaller than that of $O_1$ interstitial, which implies $O_1$ dopant is easier to be formed than $O_2$ in the bulk of $RbPbI_3$. Therefore, $RbPbI_3$ with $O_1$ interstitial is more stable than that with $O_2$ interstitial. Such observation has also been reported for other inorganic perovskite such as $CsPbI_2Br$, where $O_1$ passivated $CsPbI_2Br$ is more stable than that for $O_2$[28]. This would also benefit the formation of optimal band gaps as suggested in Table 1.

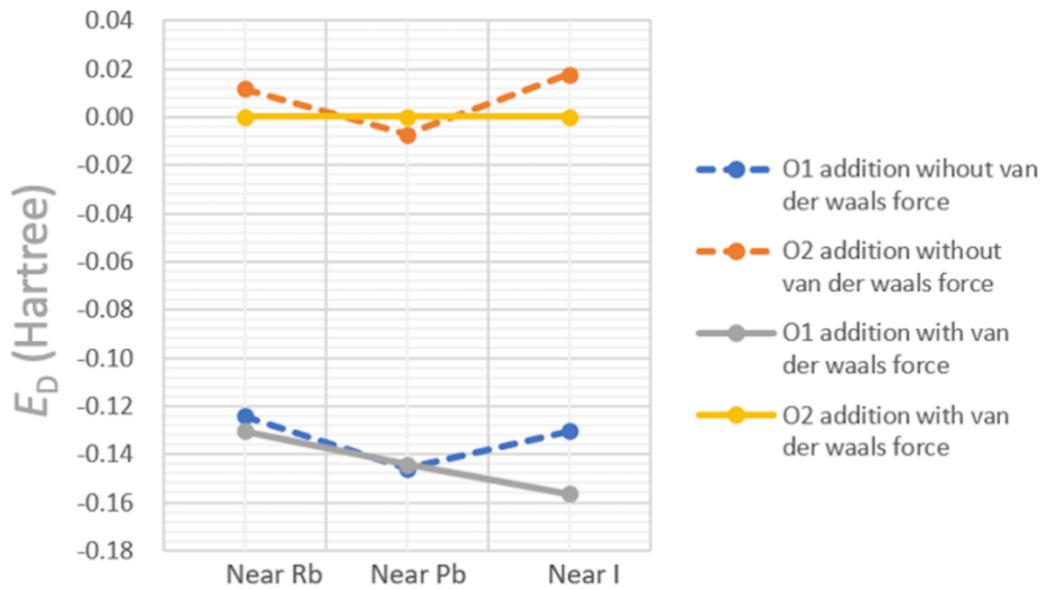

Figure 7: Defect formation energy ($E_D$) in the unit of Hartree of $O_1$ and $O_2$ interstitial, with and without vdW force.

## IV. Conclusions

First of all, the calculations with the vdW force have produced more accurate results for the band gaps as compared with experiment. Secondly, $RbPbI_3$ with $O_1$ interstitial was more stable than that with $O_2$ interstitial in terms of the defect formation energies. The crystal structure of $RbPbI_3$ remained stable after adding $O_1$ or $O_2$ in the bulk. Thirdly, oxygen interstitial would make the band gap of PSC becomes smaller down to ~0.5 eV; in the scenarios including $O_1$ interstitial and $O_2$ interstitial near-Pb, the band gap can be optimal (between 0.8 and 1.3 eV) for solar cell application. The band gap reduction is due to the formation of a defect band within the pristine band gap, which was predominated by oxygen and iodine. However, on the other hand, the narrow band width of the defect band could be problematic for electron transport. Moreover, $O_2$ interstitial could lead to an exotic metallic state, which is not desired in a solar cell device, should therefore be prevented as far as possible. However, this might be due to the high density of $O_2$ dopants. Lastly, oxygen interstitial for the bulk $RbPbI_3$ can have a significant effect on the optical properties and mobility because the defect bands could both reduce the band gap and trap electrons. But in contrast the changes in the crystal structure are not so significant as compared with those in the electronic structure. To this end, it is necessary to control oxygen dopant level appropriately in the fabrication process.

The previous results show that the passivation of oxygen atoms on the surface could improve the efficiency of PSC[28]. Our work suggested the $O_1$ interstitial could have a positive impact on the electronic and optical performance of PSC in terms of the size of the band gaps, depending on the scenarios, which is consistent with the results of oxygen passivation of $CsPbI_2Br$[28] stating that $O_1$ passivation could improve the PCE. In this report, the oxygen doping density in the bulk is in the order of $10^{21}$ per $cm^3$; this might be another reason why we can see a metallic state. Hence varying the doping concentration or the size of the supercell can be one of the future topics to explore. By contrast, to our best knowledge, the experimental data about oxygen defect densities in PSC is very rare for the moment. However, it should be noted the analysis for the effects of dopant on the other elements would be valid as the atomic distances between oxygen atoms remain large between those in the neighbouring cells. Moreover, the substitutional oxygen dopant could exist in the bulk as well, which therefore might be an interesting topic in the future study. In addition, since it is difficult to obtain perfect crystals in the actual processing of PSC, vacancy is also an important factor affecting the stability of PSC. According to previous research, the iodine vacancies can even increase the stability of PSC in the process of oxygen passivation[28]. Therefore, the study of iodine vacancies is particularly important. Furthermore, the influence of oxygen atoms and molecules on the surface of $RbPbI_3$ may be different from that in the bulk, as suggested by the previous similar studies[28], which could be studied in the future.

# DATA AVAILABILITY

All the computer codes and data that support the findings of this study are available from the corresponding author upon reasonable request.

# ACKNOWLEDGMENTS

This work was supported by the EU Horizon 2020 Project Marketplace, No. 760173. The authors would like to acknowledge the support provided by the Institute for Materials Discovery (IMD) at University College London.

# AUTHOR CONTRIBUTIONS

WW and KLC contributed to the conception of the paper. CY performed calculations under the supervision of WW. CY and WW analysed the theoretical data. All the authors wrote the paper.

# COMPETING INTERESTS

The authors declare no competing interests.

[1] S. K. Sahoo, B. Manoharan, and N. Sivakumar, "Introduction: Why perovskite and perovskite solar cells?", *Perovskite Photovoltaics: Basic to Advanced Concepts and Implementation*, Elsevier, 2018, pp. 1–24.

[2] S. K. Sahoo, "Renewable and sustainable energy reviews solar photovoltaic energy progress in India: A review," *Renewable and Sustainable Energy Reviews*, vol. 59. Elsevier Ltd, pp. 927–939, Jun. 01, 2016.